\pdfoutput=1
\documentclass[aps,twocolumn,superscriptaddress]{revtex4-2}
\usepackage[colorlinks=true,citecolor=blue,urlcolor=blue,linkcolor=blue,pdfstartview=FitH,bookmarksopen]{hyperref}
\usepackage{amsmath,amssymb}
\usepackage{bm}
\usepackage{comment}
\usepackage{graphicx}

\renewcommand\d\partial
\newcommand\+\dagger
\newcommand\x{\textbf{x}}

\begin{document}
\title{Droplets of Bosons at a Narrow Resonance}
\author{Ke Wang}
\affiliation{James Franck Institute and Kadanoff Center for Theoretical Physics, University of Chicago, Chicago, Illinois 60637, USA}
\author{Thimo Preis}
\affiliation{Institut f\"ur Theoretische Physik, Universit\"at Heidelberg, 69120 Heidelberg, Germany}
\author{Dam Thanh Son}
\affiliation{James Franck Institute and Kadanoff Center for Theoretical Physics, University of Chicago, Chicago, Illinois 60637, USA}

\date{July 2024}

\begin{abstract}
We consider bosons interacting through a narrow $s$-wave resonance.  Such a resonance is characterized by an infinite scattering length and a large and negative effective range $r_0$. We argue that any number $N\ge3$ of bosons can form a self-bound cluster with the binding energy per particle increasing as $N^2$ for $1\ll N\ll (-r_0/a_\text{bg})^{1/2}$, where $a_\text{bg}$ is the background scattering length (between atoms and molecules). In the opposite limit $N\gg (-r_0/a_\text{bg})^{1/2}$, bosons form droplets with binding energy per particle saturating to a constant value independent of the particle number. The stability of clusters and droplets when the interaction is detuned from the resonance is also studied.
\end{abstract}

\maketitle

\section{Introduction}
Dilute quantum droplets are particularly interesting physical systems that have recently
become the subject of active theoretical and experimental study~\cite{Ferrier-Barbut-PT}. The oldest known quantum droplets are those formed by $^4\text{He}$ atoms~\cite{Whaley:1994,Slenczka-Toennies}.  Such a droplet~\footnote{We use the terms ``droplet'' and ``cluster'' interchangeably.} exists for any number of helium atoms $N$, with binding energy per particle approaching the thermodynamic limit of $7.1\,\text{K}$ when $N\to\infty$.  It is notable that the approach to this asymptotic value is quite slow, and, in the intermediate regime $3\leq N\lesssim 10$, the binding energy, instead of growing linearly, follows an approximate quadratic law of $(N-2)^2$~\cite{Blume:2000}.  It was suggested that the bosons with short-range attractive interaction can form a droplet in 3D~\cite{Bulgac:2001wc} and 2D~\cite{Hammer:2004as}.  In 2015, Petrov~\cite{Petrov:2015} showed that in a bosonic mixture the collapse of the system can be prevented by quantum effects (see also Ref.~\cite{Petrov:2016}). Droplets stabilized by quantum effects have been realized experimentally~\cite{Ferrier-Barbut:2016,Cabrera:2018}. 


{ In this paper, we study the ground state of bosons that interact with each other via a narrow 
s-wave resonance, and we show that this ground state is a bound state.}  With ultracold atoms, narrow resonances are realized, for example, by cesium atoms in a magnetic field~\cite{Mark:2018}. At such resonances, bosons can form both atomic and molecular condensates; the thermodynamics and dynamics of such a hybrid condensate have been studied~\cite{Tommasini1998,timmermans1999rarified,timmermans1999feshbach}. In particular, it was found that such a system can form a ``mutually trapped state''~\cite{cusack2001existence}, in essence, a self-bound drop of a liquid phase that phase-coexists with the vacuum.

At the other end of the range of particle numbers, 
the three-body problem of bosons with narrow resonance has been solved~\cite{Nishida2012,Griesshammer:2023scn}. The properties of the three-body system is completely characterized by the scattering length $a$ and the effective range $r_0$, both assumed to be much larger than any other length scales in the problem, and, in addition, $r_0$ is assumed to be negative.  At $a=\infty$ there is an infinite tower of three-body Efimov states, with the energy of the ground state entirely determined by $r_0$. It has been found that a stable three-body bound state exists only in a finite range of inverse scattering length $1/a$: when $1/a$ is too negative, the three-body bound state decays into free particles. Conversely, when $1/a$ is too large and positive, the three-body bound state is unstable toward a decay into a particle and a bound dimer.  { In nonlinear optical media, a scenario  similar to a narrow Feshbach resonance are intensively studied, where bound state are discovered\cite{PhysRevLett.78.4749,PhysRevLett.81.3055,PhysRevA.58.2488}. }

We will show that a system of a large number of bosons at a narrow $s$-wave resonance forms a bound cluster with binding energy that first increases as $N^3$ and whose size decreases as $1/N$ with increasing $N$.  As $N$ is increased further, one goes to the regime of large droplets with constant density and the core and binding energy increase as
$\mathcal{O}(N)$ (i.e., constant binding energy per particle).  The crossover between the two regimes happens at $N\sim (r_0/a_\text{bg})^{1/2}$, where $a_\text{bg}$ is the characteristic scale of the background atom-atom, atom-molecule, and molecule-molecule scattering. Moreover, we show that the $N$-body bound state remains stable in a range of the detuning parameter $|r_0|/a$, which includes both positive and negative values of $a$.  The width of this range scales as $N^2$ in the small cluster regime and is constant in the large droplet regime. Combining our results with the solution to the three-body problem~\cite{Griesshammer:2023scn}, one comes to the conclusion that at a narrow resonance, a bound state of $N$ bosons exists for any $N\ge3$.

\section{The model}

A narrow resonance is characterized by a large $s$-wave scattering length $a$ and a large and negative effective range $r_0$~\cite{Petrov2004,gurarie2007resonantly}.  The effective Hamiltonian describing this situation is (we set the mass of the atom $m=1$, so the mass of the molecule is 2)  
\begin{multline}\label{eq:theory}  
   H = \int\!d\x \biggl[  \frac12 |\nabla\psi|^2+ 
   \frac14 |\nabla\phi|^2\\
   - \alpha(\psi^\+\psi^\+ \phi + \phi^\+ \psi\psi)
   + \nu {\phi^\+\phi} \biggr] .
\end{multline}
By computing the atom-atom scattering amplitude from~(\ref{eq:theory}), one can establish the connection between the parameters $\alpha$ and $\nu$ with the scattering length $a$ and the effective range $r_0$ characterizing the low-energy interaction between the atoms,
\begin{equation}
 \alpha = \sqrt{\frac{4\pi}{-r_0}}, \qquad 
  \nu = - \frac2{(-r_0) a}\, .
\end{equation}
In particular, for negative detuning $\nu<0$ the molecule is bound in vacuum, while for $\nu>0$ it is unbound. In the regime $a\ll r_0$, the binding energy of the dimer is $-\nu$.

As a nonrelativistic quantum field theory, the theory given by Eq.~(\ref{eq:theory}) is superrenormalizable. Indeed, the dimensions of both $\alpha$ and $\nu$ are positive: $[\alpha]=\frac12$ and $[\nu]=2$.  Thus the theory~(\ref{eq:theory}) can be defined without an ultraviolet cutoff.

In the real world (\ref{eq:theory}) is only an effective field theory. There exist irrelevant corrections to the Lagrangian with coefficients of natural (i.e., not finely tuned) magnitudes, and they limit the regime of validity of~(\ref{eq:theory}) in the ultraviolet.  If we denote by $a_\text{bg}$ the length scale associated with these terms, we can safely use Eq.~(\ref{eq:theory}) when the characteristic momentum is much smaller than $1/a_{\text{bg}}$. For now, let us assume that $a_{\text{bg}}=0$, and check the effect of the irrelevant corrections to Eq.~(\ref{eq:theory}) later.

\section{Droplets}
We will try to find the ground state of $N$ bosons, with $N\gg1$. First let us assume the resonance is at exact zero energy, i.e., the detuning parameter vanishes $\nu=0$. We expect that the mean-field approximation works for $N\gg1$ bosons.  Thus the problem becomes that of minimizing the classical energy functional given by Eq.~(\ref{eq:theory}) with $\nu=0$ under the constraint
\begin{equation}
  \int\! d\x\, (\psi^\+\psi + 2\phi^\+\phi) = N\, .
\end{equation}

There are two competing contributions to the energy of the droplet: the positive gradient energy and the negative attraction energy between the two condensates. To understand the interplay between these contributions, one can perform a simple variational calculation. Namely, we pick a profile function $f(r)$ which has a finite value at $r=0$ and tends to zero exponentially at $r=0$. For definiteness, we take
\begin{equation}\label{eq:shape}
  f(x) = \frac1{\cosh(x)}  \, .
\end{equation}
We then try the following ansatz for the condensates of atoms and molecules,
\begin{subequations}\label{eq:ansatz}
\begin{align}
  \psi(r) &= \sqrt{\frac{cN}{4\pi I_2 R^3}}\, f\left( \frac rR\right), \\
  \phi(r) &= \sqrt{\frac{(1-c)N}{8\pi I_2 R^3}} \, f\left( \frac rR\right)  .
\end{align}
\end{subequations}
Here $I_2$ is a numerical constant that depends on the shape function $f(x)$,
\begin{equation}
\label{6}
  I_2=\int_0\limits^\infty\!dx\, x^2 f^2(x)\, ,
\end{equation}
with $I_2\approx 0.822$ for the choice~(\ref{eq:shape}).
The ansatz~(\ref{eq:ansatz}) corresponds to two clouds of atoms and molecules of the same shape and size. The variational parameter $R$ controls the size of the droplet, and $c$ denotes the fraction of free atoms not bound in molecules. The fraction of atoms bound in molecules is $(1-c)$, and the total particle number is $N$. 

In addition to $m=1$ we will further set $-r_0=1$ for convenience. In particular, energy is measured in units of $\hbar^2/(mr_0^2)$. Inserting the ansatz~(\ref{eq:ansatz}) into the energy~(\ref{eq:theory}), one finds the variational energy
\begin{equation}
  E(c,R) = a(c)\frac{N}{R^2} - b(c)\frac{N^{3/2}}{R^{3/2}}\,,
\end{equation} 
with
\begin{equation}
  a(c) =  \frac{3c+1}8 \frac{K}{I_2}\,, \quad b(c) =  c\sqrt{2(1-c)} \, \frac{I_3}{I_2^{3/2}} \, ,
\end{equation}
where we have defined two further characteristics of the shape
\begin{equation}
  I_3 =\int\limits_0^\infty\! dx\, x^2 f^3(x)\, , \quad K =\int\limits_0^\infty\!dx\,x^2 (f'(x))^2\, .
\end{equation}
For the choice~(\ref{eq:shape}), $I_3\approx 0.367$ and $K\approx 0.607$. The gradient terms in the energy lead to the $1/R^2$ contribution, which dominates at small $R$, and the Feshbach interaction [the terms proportional to $\alpha$ in Eq.~(\ref{eq:theory})] leads to the $-1/R^{3/2}$ contribution, which dominates at large radii. At fixed $c$, the energy is minimized at radius
\begin{equation}
  R = \frac{16a^2}{9b^2}\frac1 N\,,
\end{equation}
with the value at the minimum
\begin{equation}
   E(c) = - \frac{27}{256} \frac{b^4}{a^3} N^3
  = -A \frac{c^4(1-c)^2}{(c+\frac13)^3} N^3 ,
\end{equation}
where $A=8I_3^4/(I_2K)^3\approx1.16$.
Note that $E_0$ is zero both at $c=0$ and $c=1$: at these values one of the condensates vanishes and the coupling term $-\phi\psi^2$ does not give an attractive contribution to the energy.  The minimal energy is achieved at the atomic fraction
\begin{equation}
      c= \frac{\sqrt{17}-1}6 \approx 0.521,
\end{equation}
with the energy at the minimum 
$E_0  \approx -0.0316 N^3$. Since this is only a variational calculation, this should be regarded as an upper bound on the energy. Indeed, a variational calculation based on the modified Woods-Saxon shape function $[\cosh(x/R)+\xi]^{-1}$, with five variational parameters corresponding to the parameters $R$ and $\xi$ of the atom and the molecular condensates, and the relative amplitude of the two condensates, yields 
\begin{equation}\label{eq:GS-energy}
  E_0=-0.0347 N^3 \frac{\hbar^2}{mr_0^2}\, ,
\end{equation}
where we have restored the unit of energy previously set to 1. Further minimization using the imaginary-time Gross-Pitaevskii equation lowers the energy by less than the last digit given in Eq.~(\ref{eq:GS-energy}) \cite{SM}.

We now consider the fate of the $N$-boson cluster when the detuning parameter $\nu$ is nonzero. It was found in Ref.~\cite{Griesshammer:2023scn} that the three-body bound state (the trimer) exists only within a finite range of $\nu$, $\nu_-(3)<\nu<\nu_+(3)$, where $\nu_+(3)\approx8.72$ and $\nu_-(3)\approx -0.366$. For $\nu<\nu_-(3)$, the trimer decays into a dimer (molecule) and an atom, while for $\nu>\nu_+(3)$ the trimer is completely unbound.  We now show that the $N$-body bound states at large $N$ behave in a similar way: each of them exists in a finite range of the detuning parameter, $\nu_-(N)<\nu<\nu_+(N)$, but the range expands with increasing $N$: $\nu_\pm(N)\sim N^2$. 

The qualitative behavior of the $N$-body cluster can again be investigated using the variational ansatz. The contribution of the detuning term to the energy is simply proportional to the number of molecules and independent of the cluster size $R$. Thus minimization over $R$ proceeds as in the case of zero detuning and one obtains the energy as a function of atomic concentration $c$,
\begin{equation}
  E = - N^3 A \frac{c^4(1-c)^2}{(c+\frac13)^3} + \frac \nu2 N (1-c)\, .
\end{equation}
The behavior of this function of $c$ is controlled by $\tilde\nu=\nu/N^2$. There are two critical values, which in our variational calculation are $\tilde\nu_+\approx0.132A\approx 0.153$ and $\tilde\nu_-\approx-0.117A\approx -0.136$.
For $\tilde\nu_-<\tilde\nu<\tilde\nu_+$, the minimum of $E(c)$ is located at a finite value of the atomic fraction $c$  which changes from $0.667$ at $\tilde\nu=\tilde\nu_+$ to $0.404$ at $\tilde\nu=\tilde\nu_-$. A more accurate treatment~\cite{SM} gives the values of the upper and lower ends of the interval, beyond which the $N$-boson bound state ceases to exist, as
\begin{equation}\label{eq:nupmN}
  \nu_+(N) \approx 0.191 N^2, \quad \nu_-(N) \approx-0.140 N^2\, .
\end{equation}

If one fixes the value of the detuning parameter $\nu$, parametrically larger than 1 (in the unit system $m=-r_0=1$), then a bound cluster can only be formed if the number of particles is larger than some $N_\text{crit}=(\nu/\tilde\nu_\pm)^{1/2}$ for positive or negative detuning, respectively.

As we have seen, the size of the cluster decreases with increasing number of particles as $1/N$. The density at the center of the cluster scales like $N^4$ at large $N$. That means that at some value of $N$ one can no longer ignore terms that were dropped when one writes down the energy functional. The leading irrelevant terms are 
\begin{equation}
  H_\text{bg} = \frac12\int\!d\x \left(g_{11}|\psi|^4 + 2g_{12}|\phi|^2|\psi|^2 + g_{22}|\phi|^4\right) ,
\end{equation}
where $g_{11}=4\pi a_{11}/m$, $g_{12}=3\pi a_{12}/m$, and $g_{22}=2\pi a_{22}/m$, with $a_{11}$, $a_{12}$, and $a_{22}$ being  the background (i.e, the nonresonant part of) atom-atom, atom-molecule, and molecule-molecule scattering length. As the density of the droplet increases with increasing $N$, one expects that the four-point interaction, if it is repulsive (i.e., $g_{11}>0$, $g_{22}>0$, and $g_{11}g_{22}-g_{12}^2>0$), should stabilize the density at some finite value. This can be seen by minimizing the energy using the symmetrized Woods-Saxon ansatz. At large $N$ the density distribution flattens out in the center. At very large $N$, the droplet has a bulk of constant density surrounded by a thin wall across which the density drops to zero. 

The situation here is simply the coexistence of two phases: the self-bound liquid and the vacuum.  The Landau free energy density at chemical potential $\mu$,
\begin{multline}\label{Landau}
  V(\psi,\phi; \mu) = - \alpha(\psi^\+\psi^\+\phi + \phi^\+\psi\psi)\\ + \frac{g_{11}}2|\psi|^4 + g_{12}|\psi|^2|\phi|^2 + \frac{g_{22}}2|\phi|^4 \\ - \mu(|\psi|^2 +2|\phi|^2)\, ,
\end{multline}
allows a first-order phase transition to occur at a negative value of $\mu$ where a nontrivial minimum of $\Omega(\psi,\phi)$ is degenerate with the trivial minimum at $\psi=\phi=0$. This critical value of $\mu$ and the values of condensates in the liquid phase are the solution to the equations $\Omega=\d_\psi\Omega(\psi_0,\phi_0)=\d_\phi\Omega(\psi_0,\phi_0)=0$. Parametrically, this occurs at 
\begin{equation}
  \mu\sim \frac EN \sim\frac{\hbar^2}{ma_\text{bg}|r_0|}\, .
\end{equation}
The binding energy per particle has the same order of magnitude as $\mu$. The density in the fluid phase is
\begin{equation}
  n \sim \frac1{4\pi a_\text{bg}^2 |r_0|}\, .
\end{equation}
At saturation density, the self-bound liquid is still dilute: $na_\text{bg}^3\ll 1$, justifying the mean-field approximation.  Also, provided that $a_\text{bg}$ sets the magnitude of the coefficients of the additional terms involving higher powers of fields and higher derivatives, these can be ignored in Eq.~(\ref{Landau}).

\begin{figure}[ht]
    \centering
\includegraphics[width=1\linewidth]{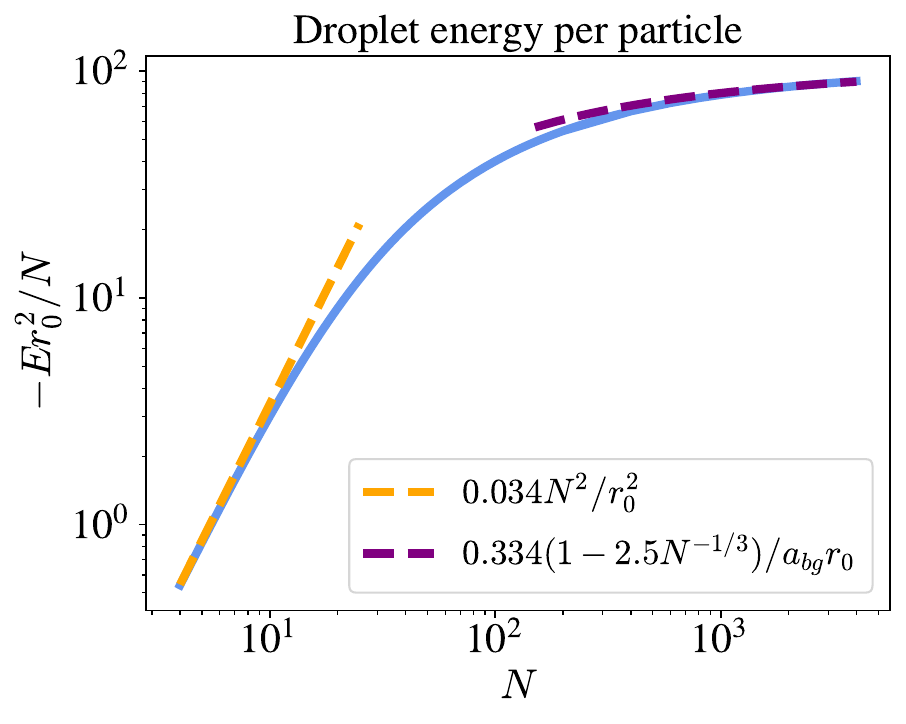}
    \caption{Droplet binding energy per particle for $g_{11}=4\pi a_{\text{bg}}$,  $g_{22}=2g_{11}/3$, $g_{12}=g_{22}/2$,  
    and $a_{\text{bg}}=|r_0|/320$. The fit at large $N$ is motivated by the picture of a droplet with finite surface tension.}   \label{droplet_energy_fig}
\end{figure}

In Fig.~\ref{droplet_energy_fig}, we plot the binding energy per particle of a droplet as a function of $N$, showing a crossover between the $\mathcal{O}(N^2)$ behavior at small $N$ to the $\mathcal{O}(N^0)$ behavior at large $N$. 

\begin{figure}[ht]
    \centering
\includegraphics[width=1\linewidth]{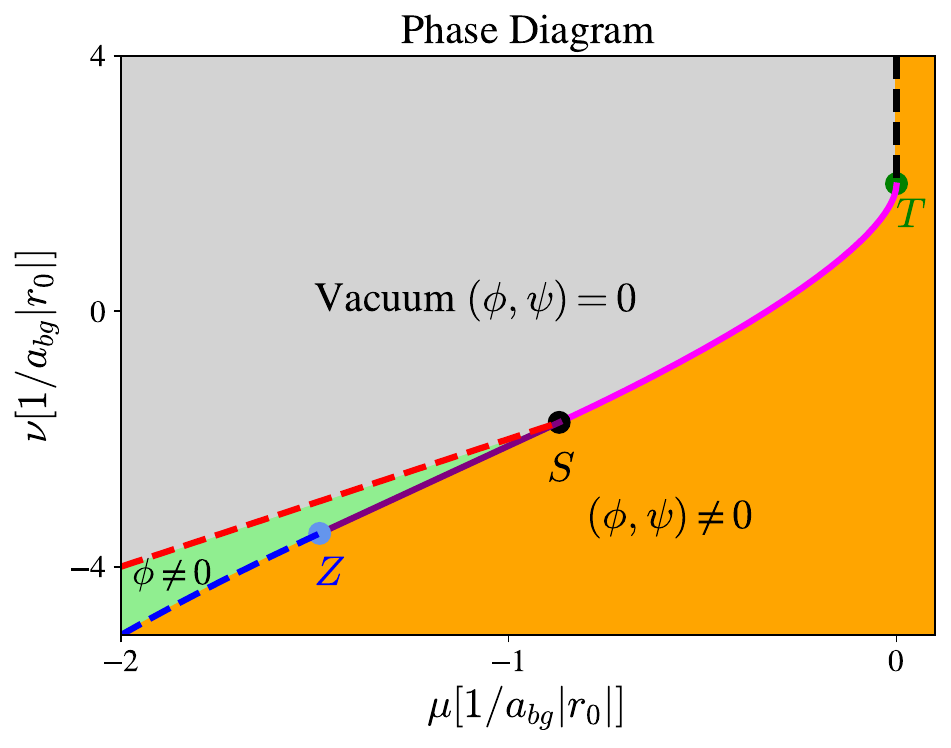}
    \caption{ 
 Phase diagram in the plane of chemical potential $\mu$ and detuning $\nu$ in units of $1/(a_{\text{bg}} |r_0|)$ for \(g_{11}=4\pi a_{\text{bg}}\), \(g_{22}=2g_{11}/3\), and \(g_{12}=g_{22}/2\).
Solid and dashed lines represent first- and second-order transitions, respectively. The grey region represents the vacuum where the particle density is zero, the green region indicates the molecular BEC phase (\(\psi=0,\phi\neq 0\)), and the  orange region corresponds to the hybrid condensate phase (\(\psi\neq 0,\phi\neq 0\)). The tricritical points are $T=(0,\nu_+(\infty))$ and $Z=(\mu_{\mathbb{Z}_2},\nu_{\mathbb{Z}_2})$~\cite{SM}, while $S=(\nu_-(\infty)/2,\nu_-(\infty))$ is the critical end point shared by the three phases.
 }   \label{droplet_phase_diagram}
\end{figure}

Now we discuss the effect of detuning.  We first consider how detuning affects the phase diagram of homogeneous matter.  For this, one needs to investigate the behavior of the Landau functional (previously considered in Refs.~\cite{timmermans1999rarified,timmermans1999feshbach,PhysRevLett.92.160402,Romans:2003acw,Lee:2004,PhysRevLett.92.050402,Radzihovsky:2008})
\begin{equation}\label{Landau-nu}
  V(\psi,\phi;\mu,\nu) 
  = 
 V(\psi,\phi;\mu) + \nu |\phi|^2\, .
\end{equation}
A typical $(\mu,\nu)$ phase diagram is shown in Fig.~\ref{droplet_phase_diagram}.  One can understand the positive detuning ($\nu>0$) region of the phase diagram by integrating out $\phi$, assuming all fields are small. The minimum of $\phi$ is achieved at
\begin{equation}
  \phi = \frac\alpha{\nu-2\mu} \psi^2 + \mathcal{O}(\psi^4) \, .
\end{equation}
Substituting this into Eq.~(\ref{Landau-nu}), we find the effective potential for $\psi$:
\begin{multline}
  V_\text{eff}(\psi) = -\mu|\psi|^2 +\biggl(\frac{g_{11}}2 - \frac{\alpha^2}{\nu{-}2\mu}\biggr)|\psi|^4 + \frac{g_{12} \alpha^2}{(\nu-2\mu)^2}|\psi|^6 \\
  + \mathcal{O}(|\psi|^8).
\end{multline}
We will limit ourselves to the case $g_{12}>0$. Then at $\mu=0$ and $\nu=\nu_+(\infty)$, with 
\begin{equation}
 \nu_+(\infty) = \frac{2\alpha^2}{g_{11}} = \frac{2\hbar^2}{ma_{11}|r_0|}\, , 
\end{equation}
the coefficients of both the $|\psi|^2$ and the $|\psi|^4$ term vanish. This point is the tricritical point $T$ in Fig.~\ref{droplet_phase_diagram}. For $\nu>\nu_+(\infty)$, as one changes $\mu$ there is a phase transition at $\mu=0$ from the vacuum to the Bose-Einstein condensate (BEC). For these values of $\nu$ the self-bound liquid does not exist. On the other hand, for $\nu<\nu_+(\infty)$ there is a first-order phase transition between the vacuum and the self-bound fluid. Thus, $\nu_+(\infty)$ is the $N\to\infty$ limit of the upper value detuning parameter for which the $N$-particle droplet exists.

At negative detuning, the line of vacuum-liquid first-order phase transition continues to exist at negative $\nu$ and meets the second-order phase transition line $\mu=\nu/2$ at a critical end point~\cite{Fisher:1991zz} (point $S$ in Fig.~\ref{droplet_phase_diagram}) at 
\begin{align}
  \nu_-(\infty) &=-\frac{2\alpha^2}{\sqrt{g_{11}g_{22}}+g_{12}}\nonumber \\
  &= -\frac{2\sqrt2\hbar^2}{m|r_0|
  (\sqrt{a_{11}a_{22}}+\frac3{\sqrt2}a_{12})}\, .
\end{align}
For $\nu$ slightly smaller than $\nu_-(\infty)$, as one increases $\mu$, the system goes through two phase transitions: first from the vacuum to a molecular BEC state with $\psi=0$, $\phi\neq0$, and then to a hybrid condensate phase with $\psi\neq0$, $\phi\neq0$. Since the hybrid condensate does not phase-coexist with the vacuum, there is no stable droplet consisting of any finite number of atoms.
The first-order phase transition line between the hybrid condensate phase and the molecular BEC phase is a straight line that terminates at a $\mathbb Z_2$ tricritical point (point $Z$ in Fig.~\ref{droplet_phase_diagram})~\cite{SM}.

\section{Conclusion}

In this paper, we have considered the problem of a droplet of a finite but large number of bosonic atoms at a narrow Feshbach resonance.  We find that the binding energy per particle should increase like $N^2$ and then flatten out to a constant.

In this work we have considered only the case when all atoms are identical bosons.  It should be straightforward to extend the calculation to the case when the narrow resonance is formed from two nonidentical bosons. 

The calculations in this paper are performed in the mean field approximation.  For a small number of particles it will be important to compute quantum corrections to the energy.  It would also be interesting if one could solve the four-body problem, e.g., by extending the zero-range calculations of Ref.~\cite{Platter:2004he} to an interaction with finite and negative effective range.  For larger $N$, one may hope to find the binding energy through Monte Carlo methods~\cite{Bazak:2018}.

Experimentally, one system where one may be able to create bound droplets is ultracold cesium at Feshbach resonance. For Cs the resonance at $B=19.849(2)\,\text{G}$ has a very small width of $\Delta=8.3(5)\,\text{mG}$~\cite{Zhang:2022fda}.  The effective range can be evaluated through~\cite{Zhai-book}
\begin{equation}
  r_{0} \simeq -\frac{2\hbar^2}{m \, \Delta \, \delta \mu\, a_{\text{bg}}}\,.
\end{equation}
Using $\delta\mu=2\pi\hbar\times0.76(3)\,\text{MHz}/\text{G}$ and $a_\text{bg}=163(1)a_0$ ($a_0$ being the Bohr radius), we find $r_0/a_\text{bg}=-320(20)$. The molecule-molecule $s$-wave scattering length is $a_{22}=220(30)a_0$~\cite{Zhang:2000}, which corresponds to $g_{22}\approx\frac23 g_{11}$.  The atom-molecule scattering is unknown, but if one assume that $g_{12}$ is of the same order of magnitude as $g_{11}$ and $g_{22}$, and that $g_{12}^2/(g_{11}g_{22})<1$, then one should expect the transition between the small cluster ($E_N\sim N^3$) regime to the large droplet ($E_N\sim N$) regime in the binding energy to occur at $N\sim (r_0/a_\text{bg})^{1/2}\sim20$.  

The authors thank Cheng Chin, Ubirajara van Kolck, Yusuke Nishida, Dmitry Petrov, Leo Radzihovsky, and Zhendong Zhang for discussion.  This work is supported, in part, by the U.S.\ DOE Grants No.\ DE-FG02-13ER41958 and by the Simons Collaboration on Ultra-Quantum Matter, which is a Grant from the Simons Foundation (No.\ 651440, DTS). T.P.\ acknowledges partial support by the DFG (EXC2181/1-390900948, 273811115) and thanks the University of Chicago for hospitality. KW acknowledges partial support from Kadanoff Center for Theoretical Physics and University of Chicago’s Research Computing Center.

\bibliography{droplet}{}
\end{document}


\onecolumngrid

\begin{center}
        \textbf{\large --- Supplemental Material ---\\ $~$ \\
        Droplets of Bosons at a Narrow Resonance}\\
        \medskip
        \text{Ke Wang, Thimo Preis, and Dam Thanh Son}
\end{center}
\setcounter{equation}{0}
\setcounter{figure}{0}
\setcounter{table}{0}
\setcounter{page}{1}
\makeatletter
\renewcommand{\thesection}{S\arabic{section}}
\renewcommand{\theequation}{S\arabic{equation}}
\renewcommand{\thefigure}{S\arabic{figure}}
\renewcommand{\bibnumfmt}[1]{[S#1]}

\section{The symmetrized Woods-Saxon variational ansatz}

We introduce the symmetrized Woods-Saxon function
\begin{equation}
   F(x;\xi,R) = \frac1{\cosh(x/R) + \xi}\,.
\end{equation}
This function goes to zero exponentially as $x\to\infty$.  It is an even function of $x$ so its derivative at $x=0$ vanishes.

We use the following five-parameter trial wave functions,
\begin{align}
  \psi(r) &= \frac{N^2}{\sqrt{4\pi I_2}} F (Nr; \xi_1, R_1), \\
  \phi(r) &= \frac{N^2b}{\sqrt{4\pi I_2}} F (Nr; \xi_2, R_2).
\end{align}
In order for the total number of particles to be equal to $N$, we require
\begin{equation}
 I_2 = \int\limits_0^\infty\!dx\,x^2\left[F^2(x;\xi_1,R_1) + 2b^2 F^2(x;\xi_2,R_2) \right] .
\end{equation}
Therefore $I_2$ is a function of the variational parameters $\xi_{1,2}$, $R_{1,2}$ and $b$.

The total energy is then
\begin{equation}\label{app:E-var}
 E = N^3 \left( \frac{K}{I_2} - \frac{2I_3}{I_2^{3/2}}\right),
\end{equation}
where
\begin{align}
  K &= \int\limits_0^\infty\!dx\, \left[\frac12 F^{\prime2}(x;\xi_1,R_1) + \frac14 F^{\prime2}(x;\xi_2,R_2) \right], \\ 
  I_3 &= b\int\limits_0^\infty\!dx\, F^2(x;\xi_1,R_1) F(x;\xi_2,R_2),
\end{align}
are both functions of the variational parameters.  We now minimize the energy~(\ref{app:E-var}) to find
\begin{equation}
   E \approx -0.0347 N^3 ,
\end{equation}
achieved at
\begin{equation}
  \xi_1\approx 0.167,\quad \xi_2 \approx 1.11, \quad R_1 \approx 1.53, \quad R_2 \approx 1.09, \quad b\approx 1.39.
\end{equation}

Now we turn on detuning.  Defining
\begin{equation}
  I_{2m} = b^2\!\int\limits_0^\infty\!dx\, x^2 F^2(x;\xi_2,R_2) ,
\end{equation}
the variational energy is now
\begin{equation}
 \frac E{N^3} = \frac{K}{I_2} - \frac{2I_3}{I_2^{3/2}} - \frac\nu{N^2} \frac{I_{2m}}{I_2}\,.
\end{equation}
For negative detuning, the droplet becomes unstable when the energy of the droplet crosses zero.  This happens at $\tilde\nu\approx -0.191$, where $\xi_1\approx-0.121$, $\xi_2\approx1.17$, $R_1\approx2.07$, $R_2\approx1.20$, $b\approx1.52$.  For positive detuning, the droplet becomes unstable when its energy is larger than the energy of $N/2$ dimers.  This happens when $\tilde\nu\approx0.140$, where $\xi_1\approx0.394$, $\xi_2\approx0.851$, $R_1\approx.143$, $R_2\approx1.22$, $b\approx1.19$.

Introducing new parameters into the ansatz (for example, by using the function $[\cosh(x/R)+\xi]^{-n}$ with $n$ being an additional variational parameter, which can have different values for atoms and molecules) does not change the ground state energy at zero detuning and the values of $\tilde\nu_\pm$ appreciably.

\section{Imaginary-time Gross-Pitaevskii equation}

To numerically find the ground state wave function of the droplet, we consider the imaginary-time Gross-Pitaevskii evolution. For simplicity, we consider the case $g_{11}=g_{22}$ with $g_{12}=0$,
\begin{eqnarray}
-\partial_\tau     \psi(r)&=&  \left( -\frac{\partial^2_r+2r^{-1} \partial_r}{2m_1}-\mu_1 \right) \psi(r)+g_{11}|\psi(r)|^2  \psi(r)-2\alpha  {\psi}^*(r) \phi(r), \\
-\partial_\tau    \phi(r)&=&  \left(-\frac{\partial^2_r+2r^{-1} \partial_r}{2m_2} -\mu_2 \right) \phi(r)+g_{22} |\phi(r)|^2  \phi(r)-\alpha  \psi(r) \psi(r).
\end{eqnarray}
The total particle number is required to be conserved during the evolution. Here, the chemical potential is evaluated self-consistently at each time step by $\mu_i=\partial E/\partial N_i$ and $N_{1/2}$ is the atomic/molecular particle number.  We place the GPE on a discrete lattice with the lattice position $r(i)=d\times i$, where $i$ the integer $0\leq i \leq L$ and $d$ is the lattice distance.

First, we use the imaginary-time Gross-Pitaevskii equation (itGPE) to confirm Eq.~(13) in the main text:  the droplet energy converges to $\approx -0.0347 N^3/r^2_0$ when $g_{11}=g_{22}=0$. We start from the symmetrized Woods-Saxon ansatz: the wave-functions are given by 
$\phi_1(r)=  \sqrt{N/4 I_2\pi  }f(r)$ and $\phi_2(r)= \sqrt{ N/4 I_2\pi  } g(r)$. Here $I_2 $ is defined in Eq.~(6) of the main text and the shape functions read
\begin{eqnarray}
f(x)=\frac{1}{\exp (x+\xi_1)/R_1+\exp -(x-\xi_1)/R_1+1}\,,\quad  g(x)=\frac{b}{\exp (x+\xi_2)/R_2+\exp -(x-\xi_2)/R_2+1} \,. \nonumber
\end{eqnarray}
The minimization of the energy functional at $N=100$ and $g_{11}=g_{22}=0$  leads to the energy $E\approx -0.034666 {N^3}/r^2_0$ with the following variational parameters: $
\xi_1\approx 0.0167r_0, R_1\approx0.01528r_0, \xi_2\approx -0.008673 r_0, R_2\approx
0.01088r_0,
b\approx0.2094$.  

Next, we simulate the itGPE with the ansatz above as the initial condition. 
\begin{figure}[h!]
    \centering
\includegraphics[width=1\linewidth]{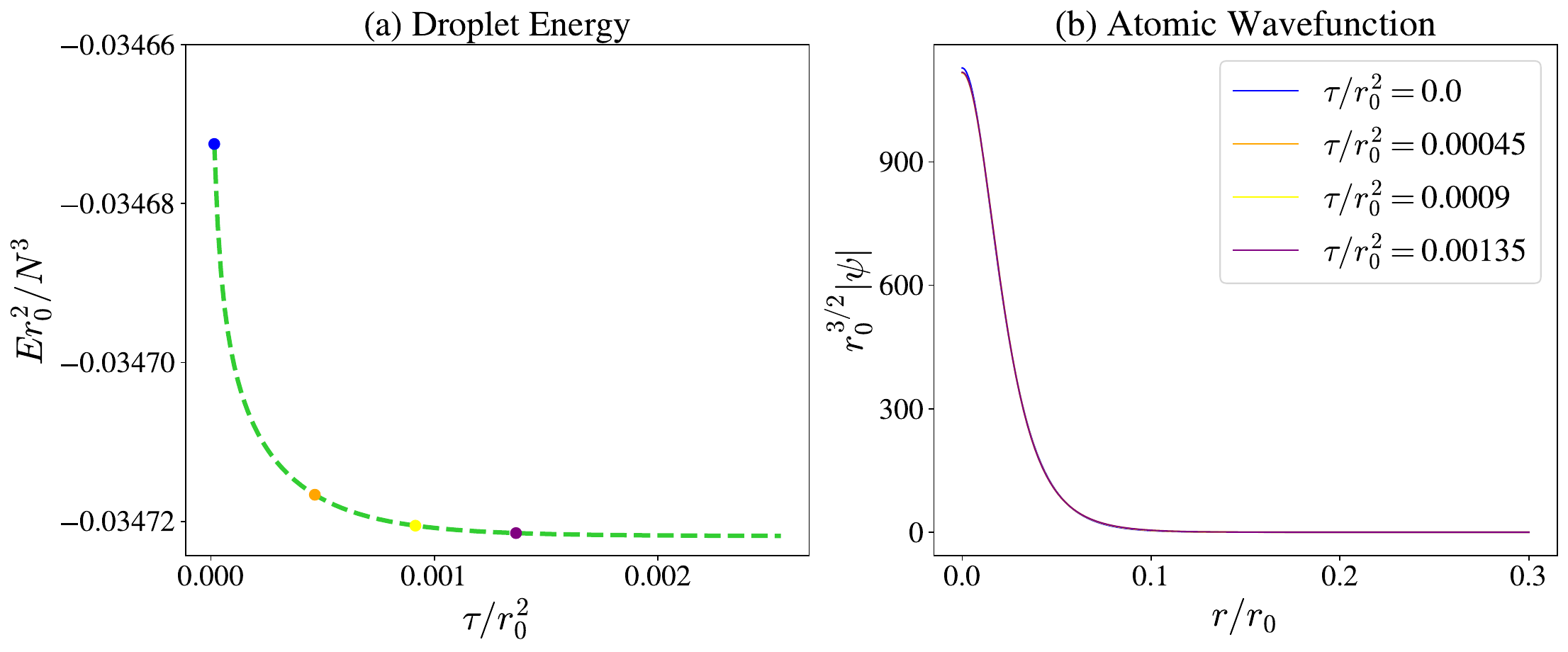}
    \caption{Imaginary time Gross-Pitaevskii evolution of the Woods-Saxon ansatz. \emph{Left}: Droplet energy versus the simulation time scale. We observe that the energy only decreases slightly and converges to $-0.03472N^3/r_0^2$. \emph{Right}: Atomic wave function at four different times. The atomic wave function changes only very slightly near $r=0$. This indicates that the Woods-Saxon ansatz is a very good approximation of the droplet's ground state.
    }
    \label{GPE0}
    \end{figure}
Here we use a lattice with $d=0.0005 r_0$ and $L=600$. Evaluation of the energy functional of the initial state {\it on} this lattice reads $\approx -0.034670 N^3/r^2_0$. Compared to the coefficient $-0.034666$ from the continuous space, one may conclude that the error from the lattice discretization on the energy of this state is of the order of $\approx 5\times  10^{-6}$ . The simulation result is shown in Fig.~\ref{GPE0}: we find that the droplet energy converges to $\approx -0.03472N^3/r_0^2$. This confirms Eq.~(13) in the main text. 

Now we aim to check the first-order transition values of $\nu_\pm(N)$ in Eq.~(15) of the main text. These two values are obtained by using the conditions $E_{\text{droplet}}(\nu_+) = 0$ and $E_{\text{droplet}}(\nu_-) = \nu N/2$. Here, $E_{\text{droplet}}$ is the ground state energy of the droplet. Using the Woods-Saxon ansatz, the conditions above lead to $\nu_+ \approx 0.191 N^2$ and $\nu_- \approx -0.140 N^2$. We pick two ansatz wave functions ($\Psi_{\pm}$), which are obtained from the minimization at $\nu/N^2 = 0.1904$ and $\nu/N^2 = -0.14$. The variational parameters for $\nu/N^2 = 0.1904$ and $N=100$ are:
$
\xi_1\approx 0.11335r_0, R_1\approx 0.019905r_0, \xi_2\approx -0.01135r_0, R_2\approx
0.01183r_0,
b\approx0.001824$.  The variational parameters for $\nu/N^2 = -0.14$ and $N=100$ are:
$
\xi_1\approx 0.003406r_0, R_1\approx 0.014305r_0, \xi_2\approx -0.00649r_0, R_2\approx
0.012214r_0,
b\approx 0.55081$.  

\begin{figure}[h!]
    \centering
\includegraphics[width=1\linewidth]{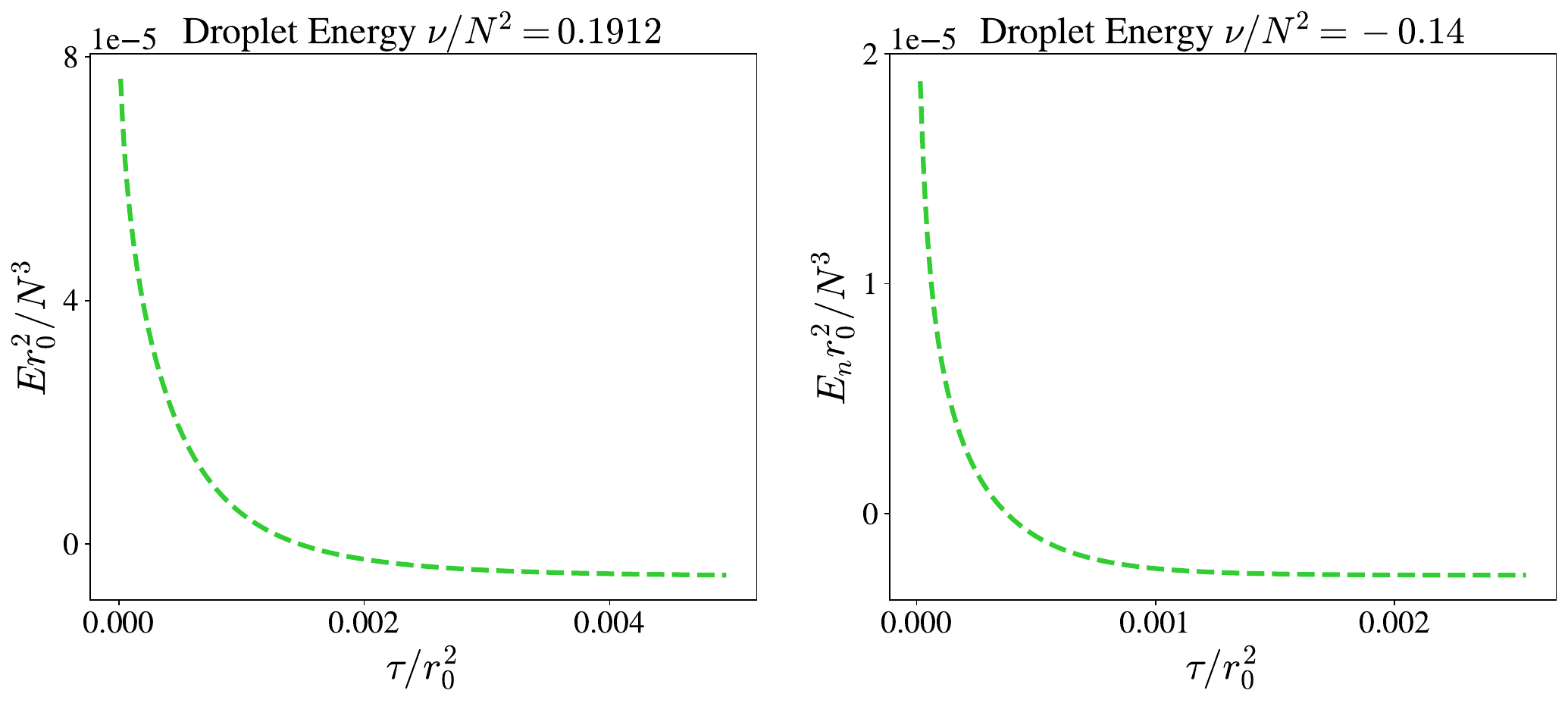}
    \caption{Imaginary time Gross-Pitaevskii evolution. \emph{Left}: Evolution of the ansatz $\Psi_+$ at $\nu = 0.1912N^2$. \emph{Right}: Evolution of the ansatz $\Psi_-$ at $\nu = -0.14N^2$. Here $E_n=E-\nu N/2$ is the energy measured from $\nu N/2$. Both energies decrease to values around zero, explicitly $\approx -5 \times 10^{-6}$ and $\approx -3 \times 10^{-6}$. 
    }
    \label{GPE1}
    \end{figure}

Then we use $\Psi_+,\Psi_-$ as the initial condition in the itGPE when $\nu/N^2$ is varied around $0.191$ and $-0.14$. The energy of the converging state represents the droplet energy. Simulation results of itGPE are discussed below ($r_0=1$).   
\begin{itemize}
\item For \(\nu_+\), we find the energy of the droplet (converging) state at \(\nu/N^2 = 0.1908, 0.1912, 0.1914\). The energy is \(E\approx -6 \times 10^{-5} N^3\) at \(\nu/N^2 = 0.1908\), while \(E\approx 2.3 \times 10^{-5} N^3\) at \(\nu/N^2 = 0.1914\). Therefore, \(\nu_+/N^2\) lies between \(0.1908\) and \(0.1914\). Furthermore, we show the imaginary time evolution in Fig.~\ref{GPE1} at \(\nu/N^2 = 0.1912\): the energy of the evolving state converges to a value around zero. Thus, we find \(\nu_+/N^2 \approx 0.191  \).
\item
For \(\nu_-\), we find the energy of the droplet state at \(\nu/N^2 = -0.1399, -0.140, -0.1403\). The  energy is \(E-\nu N/2\approx -2.4 \times 10^{-5} N^3\) at \(\nu /N^2= -0.1399 \), while  \(E-\nu N/2\approx 6 \times 10^{-5} N^3\) at \(\nu/N^2 = -0.1403\). Therefore, \(\nu_-/N^2\) lies between \(-0.1399\) and \(-0.1403\). The imaginary time evolution at \(\nu/N^2 = -0.14\) is shown in Fig.~\ref{GPE1}: the energy of the evolving state converges to a value around zero. Thus, we find \(\nu_-/N^2 \approx -0.140  \).
\end{itemize}
According to these itGPE results, we confirm the first order transition values in Eq.~(15) of the main text.

\section{The hybrid condensate-molecular BEC phase transition}

From the energy given in Eq.~(20), one finds that the phase diagram has a critical end point located at
\begin{subequations}
\begin{align}
  \mu_\text{cep} &=  -\frac{\alpha^2}{\sqrt{g_{11}g_{22}}+g_{12}}\,,\\
  \nu_\text{cep} &=  -\frac{2\alpha^2}{\sqrt{g_{11}g_{22}}+g_{12}}\,,
\end{align}
\end{subequations}
and a $\mathbb Z_2$ tricritical point located at
\begin{subequations}
\begin{align}
  \mu_{\mathbb Z_2} &= -\frac{2\alpha^2}{\sqrt{g_{11}g_{22}}+g_{12}}+ \frac{g_{12}\alpha^2}{(\sqrt{g_{11}g_{22}}+g_{12})^2} \,,\\
  \nu_{\mathbb Z_2} &= -\frac{4\alpha^2}{\sqrt{g_{11}g_{22}}+g_{12}}+ \frac{(2g_{12}-g_{22})\alpha^2}{(\sqrt{g_{11}g_{22}}+g_{12})^2} \,.
\end{align}
\end{subequations}

The line separating the molecular-BEC phase and the hybrid condensate phase is a straight line connecting $(\mu_\text{cep}, \nu_\text{cep})$ to $(\mu_{\mathbb Z_2}, \nu_{\mathbb Z_2})$:
\begin{subequations}
\begin{align}
  \mu &= -\frac{\alpha^2}{\sqrt{g_{11}g_{22}}+g_{12}} - x \frac{\sqrt{g_{11}g_{22}}\,\alpha^2}{(\sqrt{g_{11}g_{22}}+g_{12})^2}\,, \\
  \nu &= -\frac{2\alpha^2}{\sqrt{g_{11}g_{22}}+g_{12}} - x \frac{(2\sqrt{g_{11}g_{22}} + g_{22})\alpha^2}{(\sqrt{g_{11}g_{22}}+g_{12})^2}\, ,
\end{align}
\end{subequations}
where $x$ runs between 0 (the critical end point) to 1 (the $\mathbb Z_2$ tricritical point).  On the molecular-BEC side of the phase transition, the fields are
\begin{subequations}\label{eq:mBEC-cond}
\begin{align}
   \psi &= 0 ,\\
   \phi &= 
 \frac{\alpha\sqrt x}{\sqrt{g_{11}g_{22}}+g_{12}} \,,
\end{align}
\end{subequations}
and on the hybrid condensate side they are
\begin{subequations}\label{eq:hybrid-cond}
\begin{align}
  \psi &=  \left(\frac{g_{22}}{g_{11}}\right)^{1/4}\frac{\alpha\sqrt{1-x}}{\sqrt{g_{11}g_{22}}+g_{12}}\,, \\
  \phi &= \frac{\alpha}{\sqrt{g_{11}g_{22}}+g_{12}} \,.
\end{align}
\end{subequations}
One can check that (\ref{eq:mBEC-cond}) and (\ref{eq:hybrid-cond}) are two local minima of the energy and these minima are degenerate in energy.